# Sensing-Throughput analysis in NOMA-based CR Network

Deepika Rajpoot[1]

[1]Nit Kurukshetra, Electronics and Communication Engineering, Kurukshetra, Haryana
`deepikarajpoot77@gmail.com`

**Abstract.** The Cognitive Radio (CR) network provides the solution to the spectrum deficiency problem by enhancing the spectrum utilization. In recent years, Non-orthogonal multiple access (NOMA) has also gained significant interest in improving the spectrum efficiency of 5G networks. The simultaneous wireless information and power transfer (SWIPT) is a technique for the transfer of wireless information and power simultaneously for the power-limited wireless networks. In this paper, we are doing a comparative analysis between the obtainable throughput and standard throughput with perfect cancellation in the SIC receiver and also developed the golden selection search algorithm to acquire the optimal sensing time for maximizing the throughput in NOMA based CR network by using the SWIPT technique. In addition to the sensing time optimization has also simulated the throughput with the interference probability in perfect and imperfect sensing case.

**Keywords:** Cognitive Radio, NOMA, SWIPT

## 1 Introduction

As a spectrum sharing system, the CR can be utilized to enhance the spectrum utilization of wireless networks. Here, the secondary user (SU) is permitted to access the unused spectrum of the primary user (PU). But, the SU is not permitted to disrupt the ordinary PU's communication. The SU can access the allocated PU's band only when the PU is absent and as PU is detected SU has to leave the channel [1]-[2]. The PU's absence or presence is observed with the help of spectrum sensing algorithms. The energy detection is the widely used spectrum sensing algorithm due to its simple implementation. It is based on the comparison of PU's energy statistics signal with a fixed threshold [3]-[6]. The effectiveness of the above energy detection algorithm is based on false alarm probability) and the probability of detection ($P_d$) where the former represents faithful detection of an unused channel while later indicates accurate detection of PU presence. In [7], a sensing-throughput trade-off strategy was suggested to find the best sensing time to enhance the SU's throughput. For the optimal sensing duration, the SU's throughput has to be maximized under the constraint of PU protection. To fulfill the above requirement, a multiple channel CR was proposed [8], which enhances the SU's throughput by enabling the SU to access several idle subchannels simultaneously. Later on, to enhance the SU's throughput, the power optimization of multiple channels was proposed but this can be achieved only in the absence of PU [9]. In all the ideas, when PU was detected as absent, the SU still cannot utilize the spectrum. Therefore, to overcome the above drawbacks, NOMA has been proposed to enhance the spectrum efficiency of 5G communications. In the NOMA, many users can be combined on identical subchannel by involving superposition coding at the transmitter and successive interference cancellation (SIC) at the receiver [10]. Therefore, when PU is not absent, the SU can still access the spectrum to enhance the throughput by using NOMA. Also, if the non-orthogonal property of NOMA is combined with SWIPT techniques, the spectrum efficiency can be enhanced. The SWIPT enables the power's transmission and data concurrently by introducing basic changes in the receiver design. Thus, the application of NOMA and SWIPT can help in the improvement of spectrum utilization [12]. In previous work [12], the author has formulated the obtainable throughput for SU and optimized the sensing time by using the half searching algorithm.

In this paper, we have compared the standard throughput and obtainable throughput for SU and optimized the sensing time by using a golden selection search algorithm. In addition to the sensing time optimization has also simulated the throughput with the interference probability in perfect and imperfect sensing case. The rest of the paper is organized as follows: In Section 2, we present the system model of overlay CR network and the throughput of the NOMA based CR network. Section 3 describes the proposed optimization algorithm. Section 4 provides the simulation results and the conclusion is given in Section 5.

## 2 System Model

### 2.1 Overlay Model

The system model of overlay CR-NOMA network is shown fig.1. In this figure, we consider the overlay CR-NOMA network which combined of $N$ SUs, a PU and a base station (BS). In this, PU and SU harvest the energy first after this PU and SU transmit independent information to BS. Besides, SU produce the interference with PU's data transmission when spectrum sensing is performed by SU. Fig.2 shows the frame structure of overlay CR-NOMA network in which the frame is split into two distinct time slots $\tau$ and T-$\tau$. Downlink sub slot $\tau$ is utilized for SWIPT and spectrum sensing while uplink sub slot T-$\tau$ is utilized for transmitting the SU's data. During the downlink sub slot, PU and SU collect energy and PU transmits its data concurrently by using harvest-then-transmit protocol.

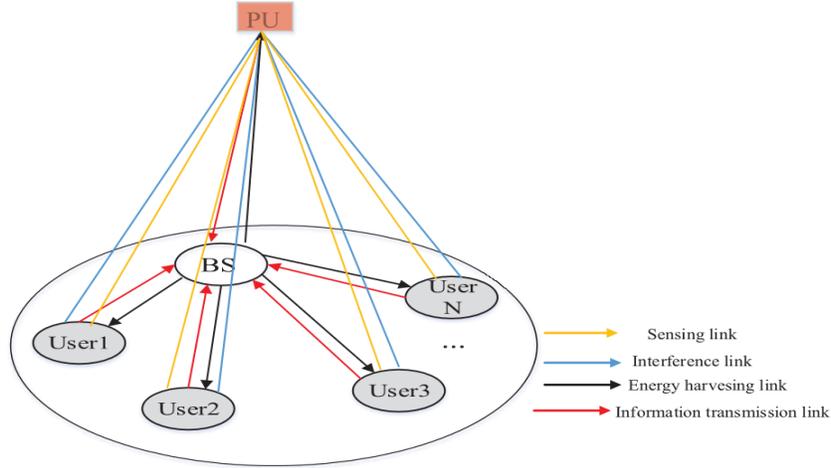

Fig.1.Network design of CR-NOMA [12].

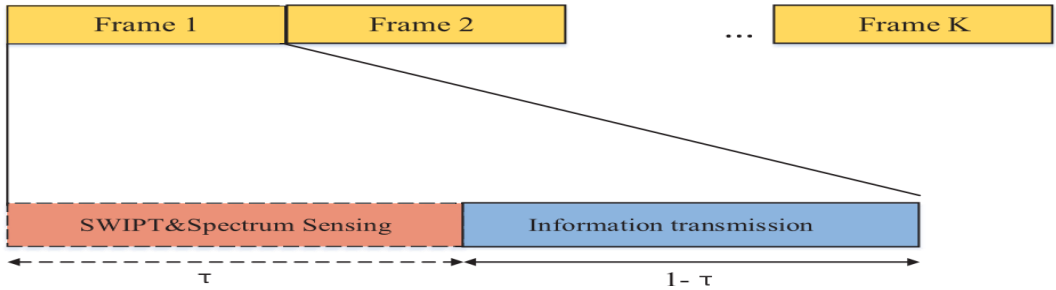

Fig.2. Frame design of CR network [12].

In [11], the author has proposed the harvest-then-transmit protocol for wireless powered communication network where BS transmit wireless energy to all users in downlink sub-slot while users broadcast their independent information to the BS in the uplink sub-slot using their independently harvested energy by Time-division-multiple-access. So, the harvested energy during the downlink sub-slot can be written as:

$$\varepsilon_i^{hd} = \tau P_{bs} \tag{1}$$

Where, $P_{bs}$ represents the BS transmit power at the transmitter. Then the transmit power for downlink sub slot can be expressed as:

$$P_T = \frac{\tau P_{bs}}{1-\tau} \tag{2}$$

This network has applied the uplink NOMA to the secondary network during the sub slot T-$\tau$. In the Uplink NOMA, more power is allocated to the nearest user while less power is allocated to the farthest user that is depend on the quality of service requirementthe hence BS employs SIC at the receiver side while all users can be combined on the identical sub channel by applying superposition coding at the transmitter side. The user who has good channel states will be decrypted first and users that are decrypted after it will be behaved as noise or interference. The $h_n$ represents user's channel gain and assume that the users are arranged as $h_1 > h_2 > \cdots\ldots h_n$ while $n(t)$ is a white gaussian noise which has zero mean and noise density $N_0(W/_{Hz})$. Therefore, the SNR and possible throughput for the n$^{th}$ user of the NOMA can be written as[11]:

$$SNR_n = \frac{P_n \gamma_n}{1 + \sum_{j=n+1}^{N} P_j \gamma_j} \quad (3)$$

$$K_n = W \sum_{n=1}^{N} log2\left(1 + \frac{P_n \gamma_n}{1 + \sum_{j=n+1}^{N} P_j \gamma_j}\right) \quad (4)$$

$$K_{ns} = W \sum_{n=1}^{N} log2\left(1 + \frac{P_n \gamma_n}{P_s g_s + \sum_{j=n+1}^{N} P_j \gamma_j}\right) \quad (5)$$

Where, $K_n$ and $K_{ns}$ is the possible throughput at the absence and presence of the PU, the normalized channel condition of the n$^{th}$ user is given by $\gamma_n = \frac{h_n}{N_0 W}$, $W$ is the transmission bandwidth of the sub band and N is the number of users. In the overlay mode, SUs essential to detect the PU's licensed spectrum intermittently. PU's radio signal is received by BS independently at the time of spectrum sensing and received signal at SU is tested for two hypotheses $H_0$ and $H_1$.

When the PU is present, the received signal can be written as:

$$H_1 : y(n) = s(n) + u(n) \quad (6)$$

When the PU is not present, the received signal can be written as:

$$H_0: y(n) = u(n) \quad (7)$$

Where $u(n)$ is represented as a white gaussian noise with zero mean and variance $\sigma_u^2$. Similarly, $s(n)$ is a random variable that is represented as PU signal with zero mean and variance $\sigma_s^2$. In the Signal to Noise Ratio (SNR), signal and noise does not depend each other. $\gamma = \frac{\sigma_s^2}{\sigma_u^2}$ is represented as SNR of PU under the hypothesis $H_1$. For deciding of the existence of PU, the *i*-th CR uses the following test statistic [11]

$$T(y) = \frac{1}{M}\sum_{i=1}^{M} |y(n)|^2 \quad (8)$$

Traditional energy detector computes the energy $T(y)$ related with received signal and corelate it with a predefined threshold ($Y_{th}$) to choose among the two hypotheses. For N number of samples, $H_1$ will be decided by Neyman-Pearson criteria if,

$$\frac{p(y/H_1)}{p(y/H_0)} > Y_{th} \quad (9)$$

where, number of samples represented by $M = \tau f_s$. We express $P_i(x)$ ($i = 0,1$) as the PDF of $T(y)$ for a given threshold. For a large $N$, a gaussian distribution with mean $\mu_1 = (\gamma + 1)\sigma_u^2$ and variance $\sigma_1^2 = \frac{1}{M}E[(|s(n)|^2 + |u(n)|^2 + s(n)u^*(n) - \sigma_s^2 - \sigma_u^2)^2]$ can approximate the $T(y)$ pdf under Hypothesis $H_1$ if $s(n)$ is complex psk modulated and $u(n)$ is circularly symmetric complex gaussian. The test statistics will be approximated as[11]:

$$\text{under } H_0, T \sim N\left(\sigma_u^2, \frac{2}{M}\sigma_u^4\right) \text{ and under } H_1, T \sim N\left((1+\gamma)\sigma_u^2, \frac{1}{M}(1+2\gamma)\sigma_u^4\right)$$

for a given threshold $Y_{th}$, The performance metrics $p_d$ and $p_f$ will be given as:

$$p_f(Y_{th}, \tau) = Q\left(\frac{Y_{th} - \sigma_u^2}{\sqrt{\frac{2}{M}\sigma_u^4}}\right), \quad p_d(Y_{th}, \tau) = Q\left(\frac{Y_{th} - (1+\gamma)\sigma_u^2}{\sqrt{\frac{1}{M}(1+2\gamma)^2\sigma_u^4}}\right) \quad (10)$$

$Q(x)$ is a complementary distribution function of standard gaussian and it is given as:

$$Q(x) = \frac{1}{\sqrt{2\pi}} \int_x^\infty exp(-\frac{t^2}{2})dt \qquad (11)$$

When we select an appropriate probability of detection ($p_d$), $p_f$ in terms of $p_d$ can be expressed as:

$$p_f = Q(\sqrt{(1+2\gamma)}Q^{-1}(p_d) + \gamma\sqrt{\tau f_s}) \qquad (12)$$

$p_d$ in terms of $p_f$ can be expressed as:

$$p_d = Q\left(\left(\frac{1}{\sqrt{1+2\gamma}}\right)(Q^{-1}(p_f) - \gamma\sqrt{\tau f_s})\right) \qquad (13)$$

### 2.2 Throughput analysis

For a given interest frequency band, let us define $P(H_1)$ is the probability of the system when PU is active and while when the PU is not active, it is denoted by $P(H_0)$. In addition, $P(H_1) + P(H_0)=1$ for conventional CR network. There are consider two possible scenarios:

- SUs generate no false alarm when PU will not present, an obtainable throughput of Noma based CR network is (1-τ) K$_n$.
- SUs does not detect the PU when it will present, an obtainable throughput of Noma based CR network is (1-τ) K$_{ns}$

The probability of these scenarios is (1-$p_f$) $P(H_0)$ and (1-$p_d$) $P(H_1)$ respectively [11]:

$$R_0(\tau) = (1-\tau)(1-p_f)\,P(H_0)K_n \text{ and } R_{0p}(\tau) = (1-\tau)(1-p_d)\,P(H_0)(1-P_p)K_n \qquad (14)$$

$$P_p = 1 - \frac{\beta}{T-\tau}\left(1 - exp\left(\frac{T-\tau}{\beta}\right)\right) \qquad (15)$$

$$R_1(\tau) = (1-\tau)(1-p_d)\,P(H_1)K_{ns} \qquad (16)$$

$$P_{ip} = \frac{\alpha}{T-\tau}\left(1 - exp\left(\frac{T-\tau}{\alpha}\right)\right) \qquad (17)$$

$$R_{1pip}(\tau) = (1-\tau)(1-p_d)(1-P_{ip})P(H_1)K_{ns} \qquad (18)$$

$P_p$ and $P_{ip}$ is the interference probability in perfect and imperfect sensing case which is explained in [12].
Then the average throughput of Noma based CR network is written as:

$$R_{th}(\tau) = R_0(\tau) + R_1(\tau) \text{ or } R_{thp}(\tau) = R_{0pp}(\tau) + R_{1pip}(\tau) \qquad (19)$$

Where, $R_{th}(\tau)$ and $R_{thp}(\tau)$ represents the standard throughput in the absence or presence of interference probability. We assume that $P(H_1)$ is small say less than 0.3, thus it is economically suitable to examine the secondary usage for that frequency band. Because K$_n$ > K$_{ns}$, therefore $R_0(\tau)$ dominates the standard throughput. Hence, the optimization throughput of secondary network can be approximated by:

$$max\,R_{th}(\tau) = R_0(\tau) \text{ or } \quad max\,R_{th}(\tau) = R_{0pp}(\tau) \qquad (20)$$

### 3 Proposed Algorithm for Sensing time Optimization

$T_{min} \leftarrow 0$ and $T_{max} \leftarrow 1$
Step-1 General purpose search technique for to find a maximum or minimum of unimodal function.
  (a) $R_0(\tau) \leftarrow (1-\tau)(1-p_f)\,P(H_0)K_n$

(b) $\tau \leftarrow$ decide the range of $\tau$ values
Step-2 Function rewritten to work in MATLAB
  (a) $f \leftarrow @(\tau) R_0(\tau)$
Step-3 To decide the lower and upper value
  (a) $\tau_{low} \leftarrow T_{min}$ and $\tau_{up} \leftarrow T_{max}$
Step-4 To compute the golden ratio and difference of upper and lower bound.
  (a) $g \leftarrow (\sqrt{5} - 1)/2$ and $d \leftarrow g \times (\tau_{up} - \tau_{low})$
  (b) $\tau_1 \leftarrow \tau_{low} + d$ and $\tau_2 \leftarrow \tau_{up} - d$
Step-5 $for\ i \leftarrow 0\ to\ 20 \leftarrow d_o$
  (a) Initialize $f(\tau_2)$ and $f(\tau_1)$
  (b) $if\ (f(\tau_2) > f(\tau_1))\ do$
  (c) $\tau_{up} \leftarrow \tau_1, \tau_1 \leftarrow \tau_2\ and\ d \leftarrow g \times (\tau_{up} - \tau_{low})$
  (d) $\tau_2 \leftarrow \tau_{up} - d\ else$
Step-6 $if\ (f(\tau_1) > f(\tau_2))\ do$
  (a) $\tau_{low} \leftarrow \tau_2, \tau_2 \leftarrow \tau_1\ and\ d \leftarrow g \times (\tau_{up} - \tau_{low})$
  (b) $\tau_2 \leftarrow \tau_{up} - d\ else$
  (c) *end if, end if* and end *for*
Step-7 $Output \leftarrow the\ optimal\ solution$
  (a) $\tau_{max} \leftarrow \tau_{up}, f(\tau)_{max} \leftarrow f(\tau_{up})$

## 4 Simulation Results

In this part, we give the performance analysis of the proposed scheme to attain the maximum throughput of the system. The comparison table shown below to explain the all throughput comparisons earlier. The analysis is done in four types of throughput i.e. obtainable throughput, standard throughput , obtainable throughput with interference probability in perfect sensing case and standard throughput with interference probability in perfect and imperfect sensing case.

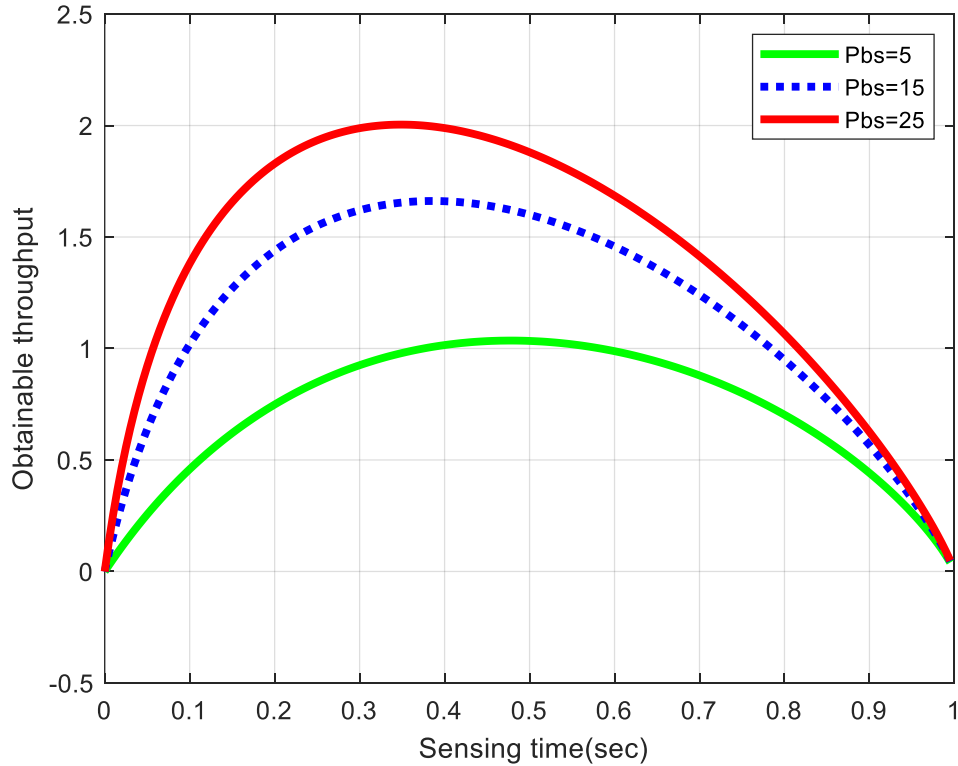

Fig.3.Optimum sensing duration ($\tau$) for maximum obtainable throughput.

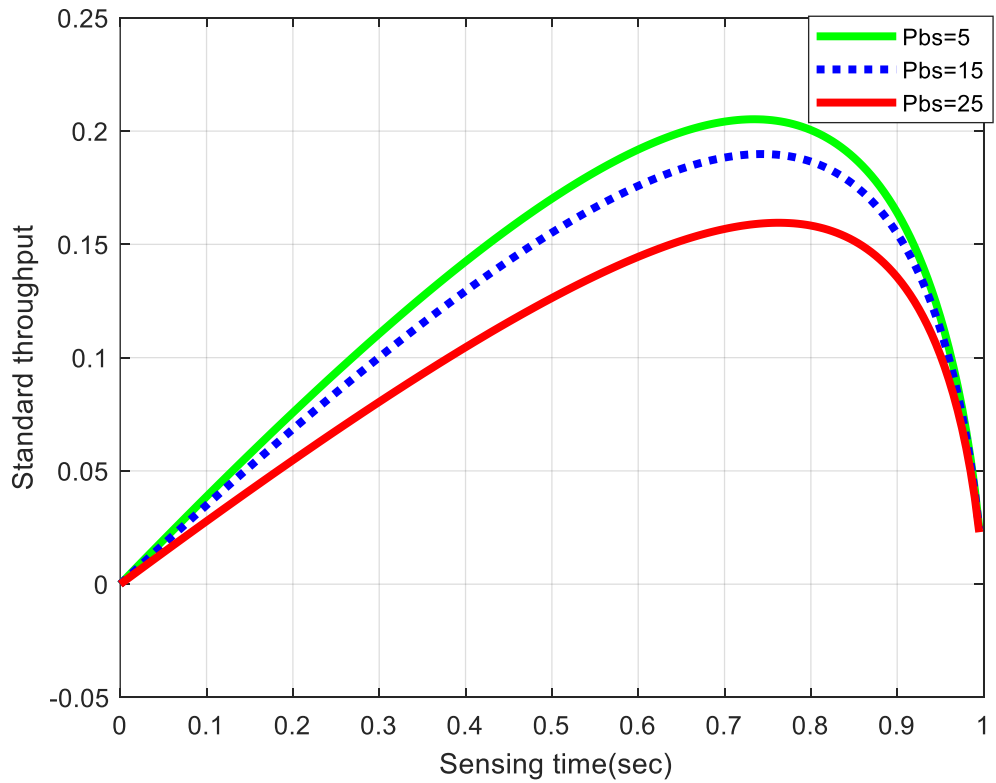

Fig.4.Optimum sensing duration ($\tau$) for maximum standard throughput.

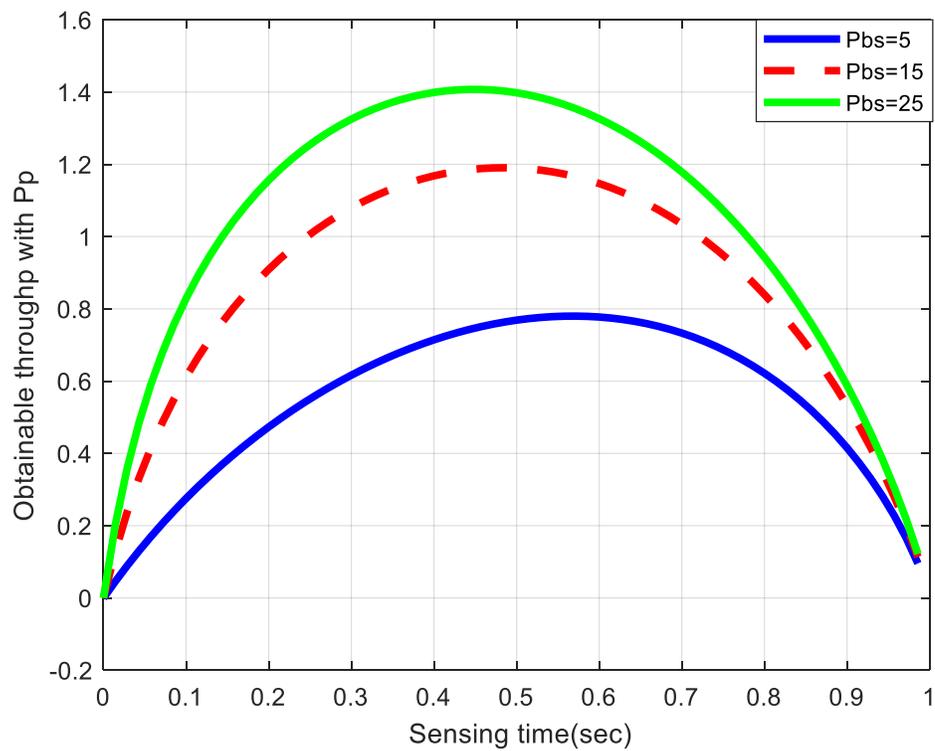

Fig.5.Optimum sensing duration ($\tau$) for maximum obtainable throughput with interference probability in perfect sensing case.

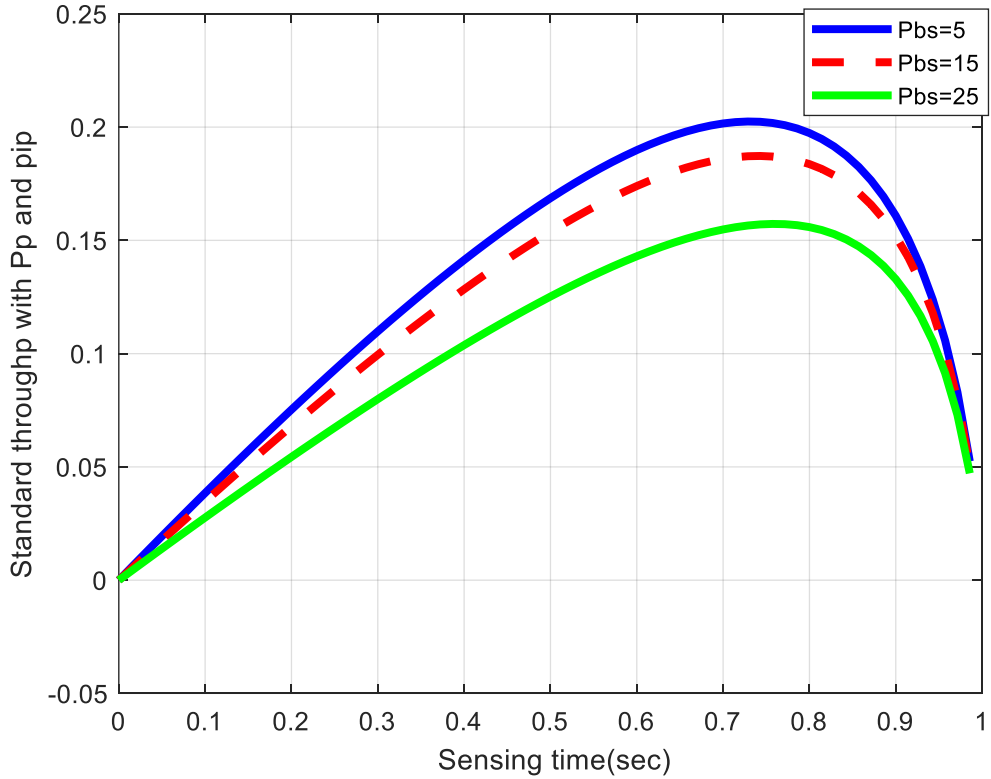

Fig.6. Optimum sensing duration ($\tau$) for maximum standard throughput with interference probability in perfect and imperfect sensing case.

Table.1 Comparison of obtainable throughput, standard throughput, obtainable throughput with interference probability in perfect sensing case and standard throughput with interference probability in perfect and imperfect sensing case

| $\tau_{min} = 0$ $\tau_{max} = 1$ | $\tau_{optm}$ (sec) | $R_0(\tau)$ (bit/hz) | $\tau_{optm}$ | $R_{th}(\tau)$ (bit/hz) | $\tau_{optm}$ (sec) | $R_{0p}(\tau)$ (bit/hz) | $\tau_{optm}$ (sec) | $R_{thp}(\tau)$ (bit/hz) |
|---|---|---|---|---|---|---|---|---|
| | 0.3438 | 2.0040 | 0.725 | 0.2052 | 0.4571 | 1.4070 | 0.7286 | 0.2025 |

From the throughput analysis, the obtainable throughput has large spectrum utilization by SU and optimum sensing time compare to standard throughput, obtainable throughput with interference probability in perfect sensing case and standard throughput with interference probability in perfect and imperfect sensing case.

## 5. Conclusion

In the sensing-throughput analysis, obtainable throughput is maximum in comparison to another throughput i.e. standard throughput, obtainable throughput with the interference probability in perfect sensing case and standard throughput with interference probability in perfect and imperfect sensing case. Standard throughput with interference probability in perfect and imperfect sensing case has less optimal sensing time and minimum throughput while obtainable throughput has more optimum sensing time and maximum throughput which maximized the spectrum utilization with the help of a NOMA-based CR network.

# References

[1] Mitola, J.: Cognitive Radio for Flexible Mobile Multimedia Communications. Mobile Netw. Appl. 6, 435- 441(2001)
[2] Ghasemi, A., Sousa, E.S.: Spectrum Sensing in Cognitive Radio Networks: Requirements, Challenges and Design Trade-offs. IEEE Commun. Mag., 46, 32-39 (2008)
[3] Shen, J., Liu, S., Wang, Y., Xie, G., Rashvand, H. F., Liu, Y.: Robust Energy Detection in Cognitive Radio. IET Commun., 3, 1016-1023, (2009)
[4] Choi, W., Song, M.G., Ahn, J., Im.G.H.: Soft Combining for Cooperative Spectrum Sensing Over Fast-Fading Channels. IEEE Commun. Lett., 18, 193-196 (2013)
[5] Liu, X., Jia, M., Tan, X.: Threshold Optimization of Cooperative Spectrum Sensing in Cognitive Radio Networks. Radio Sci.,48, 23-32, (2013)
[6] Liu, X., Jia, M.: Joint Optimal Fair Cooperative Spectrum Sensing and Transmission in Cognitive Radio. Phys. Commun., 25, 445- 453, (2017)
[7] Liang, Y.C., Zeng, Y., Peh, E.C.Y., Hoang, A.T.: Sensing-Throughput Trade-off for Cognitive Radio Networks. IEEE Trans. Wireless. Commun., 7, 1326-1337(2008)
[8] Fan, R., Jiang, H.: Optimal Multi-channel Cooperative Sensing in Cognitive Radio Networks. IEEE Trans. Wireless Commun., 9, 1128-1138, (2010)
[9] Liu, X., Li, F., Na, Z.: Optimal Resource Allocation in Simultaneous Cooperative Spectrum Sensing and Energy Harvesting for Multichannel Cognitive Radio. IEEE Access, 5, 3801-3812(2017)
[10] Higuchi, K., Benjebbour, A.: Non-orthogonal multiple access (NOMA) with Successive Interference Cancellation for Future Radio Access. IEICE Trans. Commun. 98, 403-414(2015)
[11] Song, Z., Wang, X., Liu, Y., Zhang, Z.: Joint Spectrum Resource Allocation in NOMA-based Cognitive Radio Network with SWIPT. IEEE Access, 7, 89594 – 89603, (2019)
[12] Verma, P., Singh, B.: Joint Optimization of Sensing Duration and Detection threshold for Maximizing the Spectrum Utilization. Digital Sig. Process., 74, 94-101 (2018)